\documentclass[amsmath, amssymb,10pt,aps,prb,twocolumn,notitlepage,showpacs,superscriptaddress]{revtex4-1}
\usepackage{graphicx}
\usepackage{calc}
\usepackage{braket}
\usepackage{ulem}   % to strike things out
\usepackage{slashed}
\usepackage{wasysym}
\usepackage{dsfont}
\usepackage{amsthm,amsmath,amsfonts,amssymb,verbatim,color}
\usepackage{graphicx}
\usepackage{subfigure}
\usepackage{bm}
\usepackage{epsfig,slashed}
\usepackage[T1]{fontenc}
\usepackage[colorlinks=true,citecolor=blue,linkcolor=blue,urlcolor=blue]{hyperref}
\newcommand{\Tr}{\mathop{\mathrm{Tr}}}
\newcommand{\tr}{\mathop{\mathrm{tr}}}

\renewcommand{\b}[1]{{\mathbf{#1}}}
\newcommand{\diag}{\mathrm{diag}}

\renewcommand{\c}[1]{\mathcal{#1}}

\newcommand{\Res}{\mathop{\mathrm{Res}}}

\newcommand{\Pf}{\mathop{\mathrm{Pf}}}

\begin{document}
\preprint{}

\title{Odd-frequency superconductivity in a nanowire coupled to Majorana zero modes}

\author{Shu-Ping Lee}
\affiliation{Department of Physics, University of Alberta, Edmonton, Alberta T6G 2E1, Canada}
\author{Roman M. Lutchyn}
\affiliation{Station Q, Microsoft Research, Santa Barbara, California 93106, USA}
\author{Joseph Maciejko}
\affiliation{Department of Physics, University of Alberta, Edmonton, Alberta T6G 2E1, Canada}
\affiliation{Theoretical Physics Institute, University of Alberta, Edmonton, Alberta T6G 2E1, Canada}
\affiliation{Canadian Institute for Advanced Research, Toronto, Ontario M5G 1Z8, Canada}

\date{\today}

\begin{abstract}
Odd-frequency superconductivity, originally proposed by Berezinskii in 1974, is an exotic phase of matter in which Cooper pairing between electrons is entirely dynamical in nature. The pair potential is an odd function of frequency, leading to a vanishing static superconducting order parameter and exotic types of pairing seemingly inconsistent with Fermi statistics. Motivated by recent experimental progress in the realization of Majorana zero modes in semiconducting nanowires, we show that odd-frequency superconductivity generically appears in a spin-polarized nanowire coupled to Majorana zero modes. We explicitly calculate the superfluid response and show that it is characterized by a paramagnetic Meissner effect.
\end{abstract}


\maketitle

\section{Introduction}

Superconductors can be classified by the symmetry of the pair potential, which can be thought of as the relative wave function of two electrons in a Cooper pair.  Fermi statistics require the pair potential to be odd under the exchange of those two electrons. The required sign change can come either from exchanging the opposite spins of a spin-singlet pair in even-parity (e.g., $s$-wave) superconductors, or from exchanging the opposite momenta of a spin-triplet pair in odd-parity (e.g., $p$-wave) superconductors. In 1974, Berezinskii proposed a third class of superconductors---odd-frequency superconductors---in which the pair potential satisfies the requirements of Fermi statistics not by being odd in spin space or momentum space, but by being an odd function of time or, equivalently, of frequency~\cite{Berezinskii}.

Odd-frequency pairing leads to a number of unusual features. The first is that it enables spin-triplet (singlet) pairing to appear in an $s$-wave ($p$-wave) superconductor~\cite{TedKirkpatrick1991,*TedKirkpatrick1992,BalatskyOddSC}, since antisymmetry of the pair potential under fermionic exchange is already satisfied in the time domain. The second unusual feature of odd-frequency pairing is that it leads to a vanishing static (i.e., equal-time) superconducting order parameter, because the latter is proportional to the integral over all frequencies of the pairing potential~\cite{BalatskyOddSC}. A third unusual feature predicted by theory is that odd-frequency superconductors can exhibit a paramagnetic Meissner effect~\cite{BalatskyStabilityOddSC,bergeret2001,ParamagneticMeissnerTanaka3,ParamagneticMeissnerTanaka,ParamagneticMeissnerTanaka2,
ParamagneticMeissnerJacob,fominov2015}, which may have recently been observed in experiment~\cite{ParamagneticMeissnerExp}.

Following Berezinskii's original prediction, several material platforms to engineer odd-frequency superconductors have been proposed theoretically, such as heavy fermion systems~\cite{ColemanOddSCKondo,ColemanOddSCKondo2,CoxKondoOddSC}, normal metal/superconductor junctions~\cite{tanaka2007,tanaka2007b,JacobLinder}, and ferromagnet/superconductor junctions~\cite{EfetovSFS2,EfetovOddSC,EfetovSFSRevModPhys,EfetovSFS,TanakaOddSCFMSC,TanakaOddSCFMSCReview,KeizerTriplet,TruptiTripletSC1,JuliaTriplet1,JuliaTriplet2,JuliaTriplet3}. In existing proposals odd-frequency superconductivity typically coexists with a conventional even-frequency component (unless parameters are fine-tuned), or is argued to occur in models of strongly correlated electrons where one does not have full theoretical control. Motivated by the recent experimental discovery of Majorana zero modes (MZM) in condensed matter systems~\cite{Mourik25052012,das2012zeroBiasPeak,ZeroBiasPeakInSb,FractionalJosephson,AaronZeroBiasPeak,AliYazdaniMajoranaFermion, albrecht2015}, we demonstrate via a simple exactly soluble model that coupling MZM to a spin-polarized metallic nanowire generically induces pure odd-frequency superconductivity in the nanowire without any fine tuning of parameters required~\cite{BalatskyInteractingMajorana,asano2013}.  We show by explicit calculation that the Meissner response is paramagnetic.
\begin{figure}
\includegraphics[width=7.5cm]{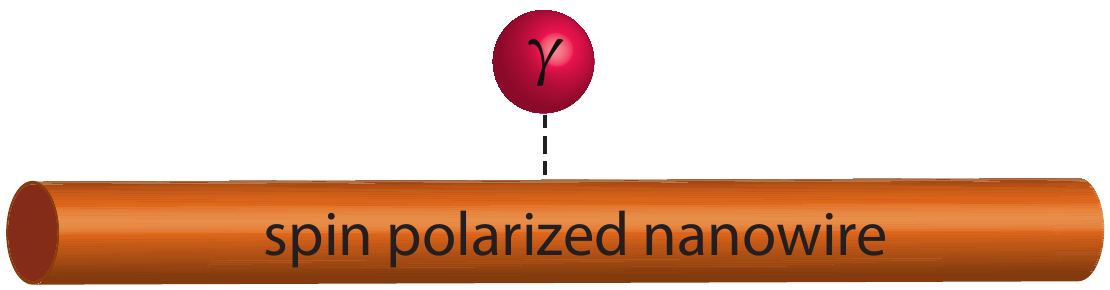}
\caption{Odd-frequency superconductivity is induced in a spin-polarized nanowire coupled to a single Majorana zero mode $\gamma$.}
\label{SingleMajorana}
\end{figure}

\section{Single Majorana zero mode}
\label{sec:singleMZM}

The fundamental building block in our proposal is the realization that odd-frequency superconductivity is generically induced in a spin-polarized nanowire coupled to a \emph{single} MZM (Fig.~\ref{SingleMajorana}). The (effectively spinless) nanowire is described by the Hamiltonian
\begin{equation}
H_w=\int dx \, c_{x}^{\dagger} \hat{\xi}(x)c_{x},
\label{SpinlessWire}
\end{equation}
where $c_x$ ($c_x^\dag$) is the annihilation (creation) operator for a spin-polarized electron at position $x$ along the wire, and $\hat{\xi}(x)$ denotes the kinetic energy operator. The coupling between the wire and a single localized MZM $\gamma_0$ at position $x=0$ can be modeled as
\begin{equation}
H_{\Gamma}^1=i\frac{\Gamma}{2}\int dx\, \delta(x)\gamma_0(c_{x}^{\dagger}+c_{x}),
\label{MajoranaCoupling}
\end{equation}
where $\Gamma$ is the coupling strength. One can rewrite the MZM in the complex fermion basis $f_0$ as $\gamma_0=f_0+f_0^{\dagger}$, and express the total Hamiltonian $H_w+H_{\Gamma}^1$ in the Nambu spinor basis $\Psi^{\dagger}(x)=(c^{\dagger}_x,c_x,f^{\dagger}_0,f_0)$ as $H_w+H_{\Gamma}^1=\frac{1}{2}\int dx \Psi^{\dagger}(x)[\hat{\mathcal{H}}_{w}(x)+M\delta(x)]\Psi(x)$, where
\begin{align}
  \hat{\mathcal{H}}_{w}(x)=
  \begin{pmatrix}
    \hat{\xi}(x) & 0 & 0 & 0 \\
    0 & -\hat{\xi}(x) & 0 & 0 \\
    0 & 0 & 0 & 0 \\
    0 & 0 & 0 & 0
  \end{pmatrix},\,
  M=\frac{\Gamma}{2}
  \begin{pmatrix}
    0 & 0 & -i & -i \\
    0 & 0 & -i & -i \\
    i & i & 0  & 0 \\
    i & i & 0  & 0
  \end{pmatrix}.
\end{align}

The easiest way to show that odd-frequency superconductivity is induced in the spin-polarized nanowire is to calculate the Green's function $G(x,y,i\omega_n)$ of the system, where $\omega_n=(2n+1)\pi T$, $n\in\mathbb{Z}$ is a fermionic Matsubara frequency and $T$ is temperature. The Green's function is given by the solution of the Dyson equation,
\begin{align}
G(x,y,i\omega_n)=G_0(x-y,i\omega_n)+ G_0(x,i\omega_n)MG(0,y,i\omega_n),
\end{align}
where
\begin{align}
&G_0(x-y,i\omega_n)=\nonumber\\
&\begin{pmatrix}
    g_0(x-y,i\omega_n) & 0 & 0 & 0 \\
    0 & -g_0(x-y,-i\omega_n) & 0 & 0 \\
    0 & 0 & 1/(i\omega_n) & 0 \\
    0 & 0 & 0 & 1/(i\omega_n)
\end{pmatrix},
\label{GreensFunction}
\end{align}
is the Green's function in the absence of the coupling (\ref{MajoranaCoupling}), and $g_0(x-y,i\omega_n)$ is the electron propagator in the wire, given by the solution of
\begin{eqnarray}
\left(i\omega_n-\hat{\xi}(x)\right)g_0(x-y,i\omega_n)=\delta(x-y).
\label{ElectronPropagators}
\end{eqnarray}
The $(2,1)$ component of the full Green's function matrix $G(x,x,i\omega_n)$ gives the induced local pairing correlator in the nanowire as
\begin{align}\label{SingleMZMPairingCorrelator}
\langle c^{\dagger}_x(i\omega_n)c^{\dagger}_x(-i\omega_n)\rangle
=\frac{-\Gamma^2g_0(x,-i\omega_n)g_0(-x,i\omega_n)}{2i\omega_n-[g_0(0,i\omega_n)-g_0(0,-i\omega_n)]\Gamma^2}.
\end{align}

Eq.~(\ref{SingleMZMPairingCorrelator}) is our first main result, and shows that the local pairing correlator is an odd function of frequency at the coupling site $x=0$. This result is completely independent of the details of the bandstructure in the nanowire. In fact, the pair amplitude at $x=0$ remains odd in frequency even if translation symmetry is broken in the wire, e.g., by disorder. If both translation and inversion symmetry are present in the wire, $g_0(x,i\omega_n)=g_0(-x,i\omega_n)$ and Eq.~(\ref{SingleMZMPairingCorrelator}) is odd in frequency for all $x$. (In the absence of inversion symmetry, the even-parity odd-frequency pairing discussed here will generically coexist with odd-parity even-frequency pairing.) A nanowire with parabolic dispersion $\hat{\xi}(x)=-\partial^2_x/(2m)-\varepsilon_F$ with $m$ the effective mass and $\varepsilon_F$ the Fermi energy gives
\begin{align}
\langle c^{\dagger}_x(i\omega_n)c^{\dagger}_x(-i\omega_n)\rangle\approx \frac{i}{2v_F}\text{sgn}(\omega_n)e^{-4|\omega_n||x|/v_F},
\label{PairingCorrelatorApprox}
\end{align}
in the low-frequency, weak-coupling limit $|\omega_n|\ll\Gamma\ll \varepsilon_F$, where $v_F$ is the Fermi velocity (see Appendix~\ref{sec:single}). The pair amplitude remains odd in frequency but decays exponentially away from the coupling site with a decay length $\sim 1/|\omega_n|$.

\section{Array of Majorana zero modes}

\begin{figure}
\includegraphics[width=8.5cm]{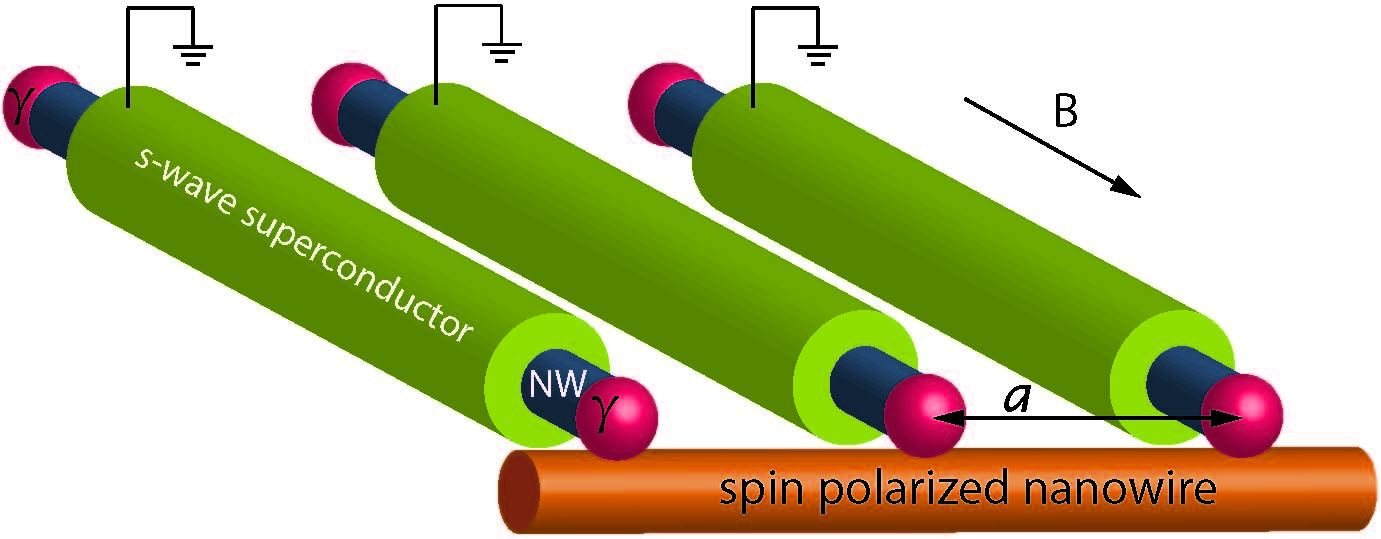}
\caption{Inducing 1D odd-frequency superconductivity in a spin-polarized nanowire. The Majorana zero modes (red dots) appear as the result of coating a regular array of strongly spin-orbit coupled semiconductor nanowires (blue wires) with conventional $s$-wave superconductors (green cylinders)~\cite{albrecht2015}, in a magnetic field $B$ parallel to the wires. The Majorana zero modes are then coupled to a spin-polarized nanowire (orange wire).}
\label{MajoranaNanoWire}
\end{figure}

Although odd-frequency pairs can be induced locally in the nanowire by a single MZM, one can in principle go a step further and engineer an extended (i.e., 1D) odd-frequency superconductor by coupling an array of MZM to the nanowire (Fig.~\ref{MajoranaNanoWire}). We consider a periodic array of strongly spin-orbit coupled nanowires coated with $s$-wave superconductors~\cite{albrecht2015}. A magnetic field is applied, which via the Zeeman effect turns these nanowires into effectively spinless $p$-wave superconducting wires~\cite{JaySau1DSemiconductor,Gil1DSemiconductor} that support unpaired MZM at their ends~\cite{Kitaev-Unpaired-Majorana-1D-wire}. Denoting the separation between two neighboring MZM as $a$, the extension of Eq.~(\ref{MajoranaCoupling}) to an array of MZM is
\begin{align}
H_{\Gamma}&=i\frac{\Gamma}{2}\sum_{n}\int dx\,\delta(x-na)(f_{x}+f^{\dagger}_{x})(c_{x}^{\dagger}+c_{x})\nonumber\\
&=i\frac{\Gamma}{2}\sum_{n}\int \frac{dp}{2\pi} (f_{p}+f^{\dagger}_{-p})(c_{p-2\pi n/a}^{\dagger}+c_{-p+2\pi n/a}),
\label{NanowireMajoranaCoupling3}
\end{align}
where we have Fourier transformed to momentum space. The problem becomes analogous to that of electrons in a periodic potential: momentum space is divided in periodic Brillouin zones of width $2\pi/a$, and the coupling between $f_{p}$ and $c_{p-2\pi n/a}$ opens a gap at the zone boundaries. We now consider a simple model problem when $a$ is much smaller than the Fermi wavelength $\lambda_F$, which captures the essence of the physics. In Appendix~\ref{sec:continuum}, we argue that odd-frequency superconductivity survives for an arbitrary MZM separation $a$.

When $a \ll \lambda_F$, the coupling between $f_{p}$ and $c_{p-2\pi n/a}$ for $n\neq 0$ only affects modes far from the Fermi energy, because the Brillouin zone edge momentum $\pi/a$ is much greater than the Fermi momentum. Thus at low energies we only need to keep the coupling to the $n=0$ mode,
\begin{align}
H_\Gamma^{n=0}=i\frac{\Gamma}{2}\int\frac{dp}{2\pi}(f_p+f_{-p}^\dag)(c_p^\dag+c_{-p}),
\label{NanowireMajoranaCouplingTrun}
\end{align}
where $p$ is restricted to the first Brillouin zone $(-\frac{\pi}{a},\frac{\pi}{a}]$. Eq.~(\ref{NanowireMajoranaCouplingTrun}) is equivalent to modeling the spin-polarized wire as a tight-binding chain of lattice constant $a$, with every site coupled to a MZM.

The superconducting wires in recent experiments~\cite{albrecht2015} have a finite length, which leads to the hybridization of the MZMs localized at the opposite ends of the wire. We model this hybridization with an energy splitting $\delta$ near the Fermi energy,
\begin{align}
H_\delta=2\delta \int\frac{dp}{2\pi}(f_p^\dag f_p-\textstyle\frac{1}{2}).
\end{align}
The combined Hamiltonian $H=H_w+H_{\Gamma}^{n=0}+H_\delta$ can be written entirely in momentum space and expressed in the Nambu basis $\Psi(p)=(c_p,c^{\dagger}_{-p},f_p,f^{\dagger}_{-p})^T$
as $H=\frac{1}{2}\int\frac{dp}{2\pi}\Psi^{\dagger}(p)\mathcal{H}(p)\Psi(p)$, where the Bogoliubov-de Gennes (BdG) Hamiltonian matrix $\mathcal{H}(p)$ is
\begin{equation}
\mathcal{H}(p)=
\begin{pmatrix}
    \xi_{p} & 0 & -i\Gamma/2 & -i\Gamma/2 \\
    0 & -\xi_{-p} & -i\Gamma/2 & -i\Gamma/2 \\
    i\Gamma/2 & i\Gamma/2 & \delta & 0 \\
    i\Gamma/2 & i\Gamma/2 & 0 & -\delta
\end{pmatrix}.
\label{BdGHamiltonian}
\end{equation}
Here $\xi_p$ is the energy-momentum dispersion of the spin-polarized wire that corresponds to the Fourier transform of $\hat{\xi}(x)$. As already mentioned, to maintain a sharp distinction between odd-frequency and even-frequency pairing in a spinless wire one must require inversion symmetry; we thus assume $\xi_{-p}=\xi_p$. We now show that the spin-polarized nanowire is a uniform odd-frequency superconductor by computing the Nambu Green's function $\mathcal{G}(p,i\omega_{n})$ as
\begin{equation}
\mathcal{G}(p,i\omega_{n})=(i\omega_n-\mathcal{H}(p))^{-1}.
\label{GreensFunctionFourier}
\end{equation}
The pair potential for electrons in the spin-polarized nanowire is obtained from the $(2,1)$ component of $\mathcal{G}(p,i\omega_{n})$ as
\begin{equation}
\langle c^{\dagger}_p(i\omega_n)c^{\dagger}_{-p}(-i\omega_n)\rangle=\frac{\Gamma^2i\omega_n/2}{\xi_p^2\delta^2+(\xi_p^2+\Gamma^2+\delta^2)\omega_n^2+\omega_n^4},
\label{PairingCorrelator}
\end{equation}
which is an odd function of $\omega_n$. Equation~(\ref{PairingCorrelator}) demonstrates that by coupling to the MZM, the spin-polarized wire effectively becomes an $s$-wave spinless odd-frequency superconductor for generic values of the couplings $\Gamma,\delta$ and for a generic inversion-symmetric normal-state dispersion $\xi_p$ in the wire. Although we have assumed translation symmetry in our derivation so far, our calculations suggest the induced odd-frequency pairing is robust against disorder in the MZM coupling $\Gamma$ (see Appendix~\ref{sec:disorder}), as one expects for on-site ($s$-wave) pairing.
\begin{figure}
\includegraphics[width=8.5cm]{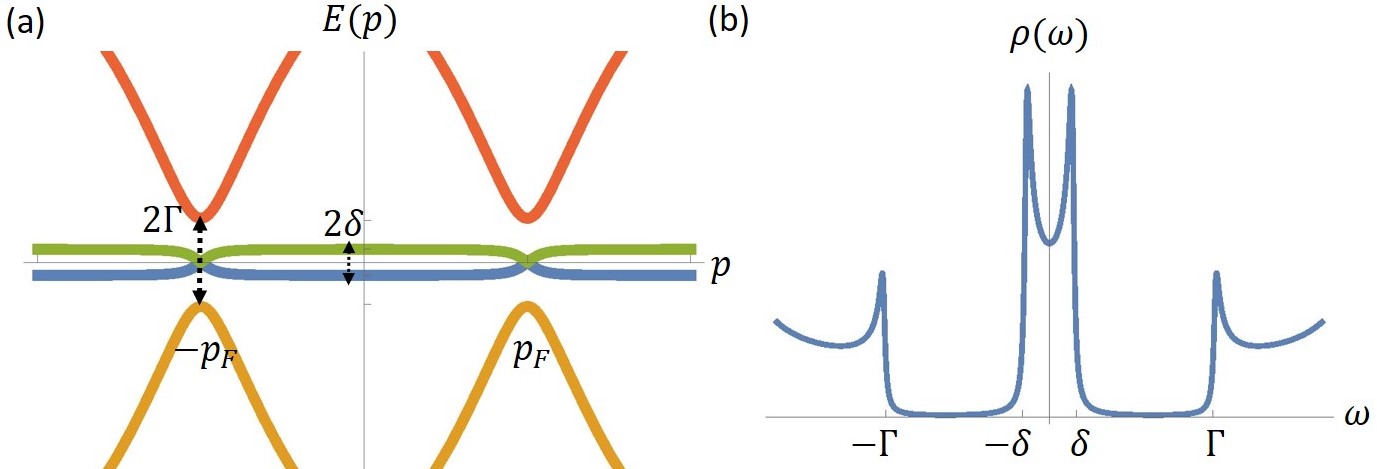}
\caption{(a) Low-energy spectrum of Bogoliubov quasiparticles in the odd-frequency superconductor; (b) electronic density of states.}
\label{BandStructureDOS}
\end{figure}

The Bogoliubov quasiparticle spectrum is obtained by diagonalizing the BdG Hamiltonian matrix (\ref{BdGHamiltonian}), and is shown schematically in Fig.~\ref{BandStructureDOS}(a) for energies near the Fermi level.  For simplicity, we consider the limit $\delta\ll\Gamma\ll\varepsilon_F$, where $\varepsilon_F$ is the Fermi energy of the spin-polarized nanowire. The Majorana modes give two nearly flat bands that become gapless at the Fermi points $\pm p_F$ due to the coupling to the spin polarized wire. In Fig.~\ref{BandStructureDOS}(b) we show a schematic plot of the electronic density of states near the Fermi level. By contrast with fully gapped even-frequency superconductors, the MZM-induced odd-frequency superconductor has a nonzero density of states at the Fermi energy which, as will be seen in the next section, leads to a paramagnetic Meissner effect~\cite{BalatskyStabilityOddSC,bergeret2001,ParamagneticMeissnerTanaka3,ParamagneticMeissnerTanaka,ParamagneticMeissnerTanaka2,
ParamagneticMeissnerJacob,fominov2015}.

\section{Meissner response}

\begin{figure}
\includegraphics[width=8.5cm]{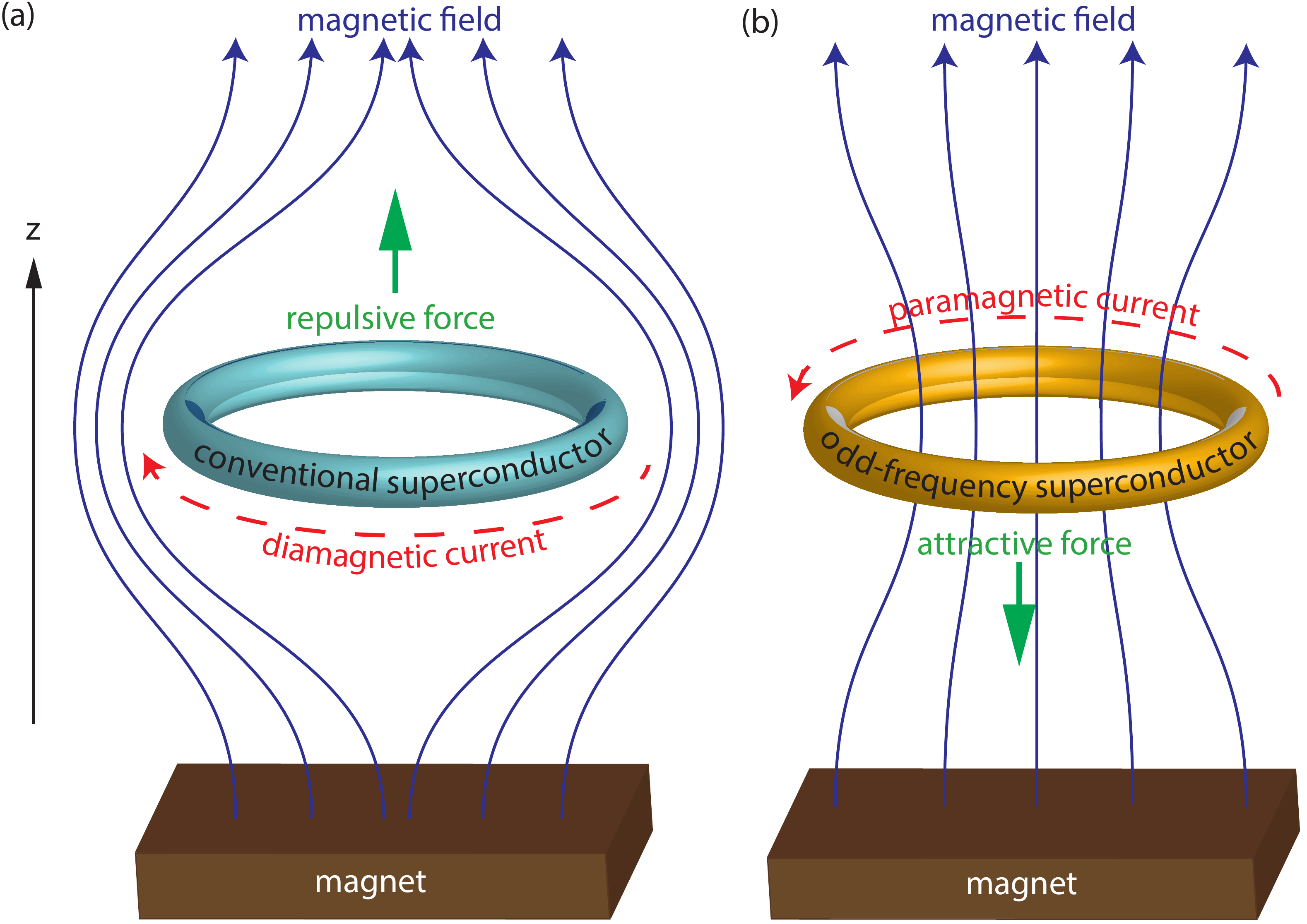}
\caption{Meissner effect of (a) a conventional superconducting ring and (b) an odd-frequency superconducting ring. In a conventional superconductor, the external magnetic field induces a diamagnetic supercurrent, which leads to a repulsive force between the ring and the magnet (superconducting levitation). By contrast, the paramagnetic supercurrent in the odd-frequency superconductor leads to an attractive force between the ring and the magnet (superconducting anti-levitation).}
\label{Meissner2}
\end{figure}

We now turn to the Meissner response of our odd-frequency superconductor. To have a well-defined Meissner effect in 1D, we consider fashioning the spin-polarized wire into a ring (Fig.~\ref{Meissner2}). A static magnetic field is applied perpendicular to the plane of the ring; this corresponds to a flux threading the ring that can be represented by a vector potential $A_x$ where $x$ is the coordinate along the ring. The Meissner response is given by the London equation
\begin{align}\label{london}
j_x=-\frac{n_se^2}{m}A_x,
\end{align}
where $j_x$ is the electric current along the ring, $e$ and $m=(d^2\xi_p/dp^2)^{-1}|_{p=\pm p_F}$ are the electron charge and mass, respectively, and $n_s$ is the superfluid density. The superfluid density for our odd-frequency superconductor can be calculated explicitly from the Nambu Green's function (\ref{GreensFunctionFourier}) following a standard diagrammatic procedure~\cite{bruus2004many,Victor,RomanGaugeInvariant}. A detailed calculation is presented in Appendix~\ref{APP:ElectromagneticResponseGeneralFormalism}, \ref{APP:DiamagneticCurrent} and \ref{APP:ParamagneticCurrent}; here we outline the main steps. The total electric current $j_x=j_x^\text{dia}+j_x^\text{para}$ is the sum of diamagnetic and paramagnetic contributions. For simplicity we compute these contributions in the limit $\delta\ll\Gamma\ll\varepsilon_F$ and $T\ll\Gamma$.

In the zero-temperature limit, the diamagnetic current is given by (see Appendix~\ref{APP:DiamagneticCurrent})
\begin{align}
j_x^\text{dia}\approx-n\left(1+\frac{\Gamma}{2\varepsilon_F}\right)\frac{e^2A_x}{m},
\end{align}
where $n=p_F/\pi$ is the electron density of the decoupled spin-polarized wire. In the presence of the coupling $\Gamma$ between the spin-polarized wire and the MZM, the diamagnetic response is simply that of the decoupled wire~\cite{bruus2004many} plus a small correction of order $\Gamma/\varepsilon_F\ll 1$. This is to be expected since the total number of electrons in the wire is not conserved in the presence of the coupling (\ref{MajoranaCoupling}) to the MZM. For the paramagnetic current, results differ depending on whether one is in the $T\ll\delta$ limit or the $\delta\ll T$ limit (see Appendix~\ref{APP:ParamagneticCurrent}). In the $T\ll\delta$ limit, we have
\begin{align}\label{JparaTlldelta}
j_x^\text{para}\approx\frac{ne^2A_x}{m}\left\{1+\frac{\Gamma}{4\delta}\left[1+\ln\left(\frac{\delta}{T}\right)\right]\right\},
\end{align}
while in the $\delta\ll T$ limit, we have
\begin{align}\label{JparaTggdelta}
j_x^\text{para}\approx\frac{ne^2A_x}{m}\left(1+\frac{\Gamma}{4T}\right).
\end{align}
Apart from the logarithmic term in Eq.~(\ref{JparaTlldelta}), the two small energy scales $T$ and $\delta$ act as an infrared cutoff for each other.

Adding the diamagnetic and paramagnetic contributions, we see that when the spin-polarized wire and the MZM are decoupled ($\Gamma=0$) both contributions exactly cancel each other and the superfluid density of the wire is zero, as expected. For $\Gamma\neq 0$, the paramagnetic contribution to the superfluid density is proportional to either $\Gamma/\delta$ or $\Gamma/T$, which is much greater than unity in the limit considered, while the diamagnetic contribution is proportional to $\Gamma/\varepsilon_F$, which is much less than unity. As a result, the paramagnetic response overwhelms the diamagnetic response, and the superfluid density is (see Appendix \ref{APP:SuperfluidDensity})
\begin{align}
\frac{n_s}{n}\approx-\frac{\Gamma}{4T}\times\left\{
\begin{array}{cc}
\displaystyle\frac{T}{\delta}\left[1+\ln\left(\frac{\delta}{T}\right)\right], & T\ll\delta, \\
\displaystyle 1, & \delta\ll T,
\end{array}
\right.
\end{align}
where we have neglected the small diamagnetic contribution proportional to $\Gamma/\varepsilon_F$. The superfluid density of the spin-polarized wire is thus negative, which is a hallmark of odd-frequency superconductivity. The dominance of the paramagnetic contribution can be traced back to the presence of gapless Bogoliubov quasiparticles at the Fermi points in the odd-frequency superconducting state. In the limit $\delta\ll\Gamma$, the density of states at the Fermi level diverges [see Fig.~\ref{BandStructureDOS}(b)] and the paramagnetic response dominates. Furthermore, the superfluid density diverges at low temperatures~\cite{YasuhiroParamagnetic,*YasuhiroParamagnetic2}, in stark contrast with its smooth behavior as $T\rightarrow 0$ in conventional superconductors~\cite{Schrieffer}.

A negative superfluid density implies a paramagnetic Meissner effect~\cite{BalatskyStabilityOddSC,bergeret2001,ParamagneticMeissnerTanaka3,ParamagneticMeissnerTanaka,ParamagneticMeissnerTanaka2,
ParamagneticMeissnerJacob,fominov2015}, whereby an applied magnetic flux is enhanced rather than screened by the induced supercurrent. In the setup of Fig.~\ref{Meissner2}, the paramagnetic supercurrent leads to an attractive force between the odd-frequency superconducting ring and a magnet [Fig.~\ref{Meissner2}(b)]. This is in sharp contrast with the repulsive force between a conventional superconductor and a magnet [Fig.~\ref{Meissner2}(a)], which leads to the phenomenon of superconducting levitation~\cite{ma2003}. To see this, we treat both the magnet and superconducting ring in Fig.~\ref{Meissner2} as magnetic dipoles with dipole moment $\b{m}_m=m_m\hat{\b{z}}$ and $\b{m}_s=m_s\hat{\b{z}}$ respectively, the latter being given by the induced supercurrent times the area of the ring. The dipole-dipole interaction produces a force on the ring given approximately by $\b{F}=-3\mu_0m_mm_s\hat{\b{z}}/(2\pi z^4)$ where $z$ is the distance between magnet and ring (assuming it is much greater than the ring radius) and $\mu_0$ is the vacuum permeability. Therefore, the superconducting ring feels an attractive force towards the magnet, a phenomenon one could call superconducting anti-levitation.

We now give a rough estimate of the magnitude of the paramagnetic Meissner current (\ref{london}). For an Al-coated InAs nanowire~\cite{albrecht2015}, the induced topological superconducting gap is $\Delta\approx 2.3$~K. For a wire length of 1~$\mu$m, this yields a MZM splitting $\delta\approx 0.1$~K~\cite{albrecht2015}. The spin-polarized nanowire can be made of ferromagnetic metals such as Co, Fe, or Ni. The coupling $\Gamma$ between the MZM and the ferromagnetic nanowire depends on the details of the sample. Since MZMs are localized at the ends of the wire, the broadening due to normal lead coupling $\Gamma$ is given by $\Gamma \sim g \Delta$~\cite{vanHeck'16}. Here $g$ is the dimensionless normal-state conductance of the contact. Assuming $g \lesssim 1$, we estimate an upper bound for the Majorana coupling as $\Gamma\sim 0.1$~K. The Fermi velocity in a Co ferromagnetic nanowire is $v_F\approx 10^{6}$~m/s~\cite{wang2010}, which we can use to extract the ratio of electron density $n$ to mass $m$ as $n/m=v_F/{(\pi\hbar)}$. Threading one flux quantum $\Phi_0$ into a ring of circumference $L\approx 1$~$\mu$m gives the electromagnetic vector potential $A_x=\Phi_0/L$ on the ring. Using these values, we find approximate upper bounds for the paramagnetic Meissner current as $j_x\approx 100$~nA at temperature $T=100$~mK in the $T\ll\delta$ regime and $j_x\approx 10$~nA at temperature $T=0.5$~K in the $\delta\ll T$ regime. We believe such currents are within experimental measurement capabilities.

\acknowledgments{
We thank Debaleena Nandi for illuminating discussions. This research was supported by NSERC grant \#RGPIN-2014-4608, the Canada Research Chair Program (CRC), the Canadian Institute for Advanced Research (CIFAR), Microsoft Station Q, and the University of Alberta. RL acknowledges the hospitality of the Aspen Center for Physics supported by NSF Grant \#1066293. JM acknowledges the hospitality of the Kavli Institute for Theoretical Physics, supported in part by NSF Grant PHY11-25915.}

\appendix

\section{Nanowire coupled to a single Majorana zero mode}
\label{sec:single}

Here we provide a detailed derivation of the results presented in Sec.~\ref{sec:singleMZM}. We start from a spin-polarized nanowire coupled to a single Majorana zero mode as shown in Fig.~\ref{SingleMajorana}. For simplicity, we model the Hamiltonian of the (effectively spinless) nanowire as
\begin{equation}
H_w=\int dx \, c_{x}^{\dagger}\left(-\frac{\partial_x^2}{2m}-\varepsilon_F\right)c_{x},
\label{SpinlessWire}
\end{equation}
where $c_x$ ($c_x^\dag$) is the annihilation (creation) electron operator in continuous space. The coupling of a single Majorana mode can be modeled as
\begin{equation}
H_{\Gamma1}=i\frac{\Gamma}{2}\int dx\, \delta(x)\gamma_0(c_{x}^{\dagger}+c_{x}).
\label{MajoranaCoupling2}
\end{equation}
Here we assume the Majorana zero mode $\gamma_0$ couples to the nanowire at $x=0$ and $\delta(x)$ is the Dirac delta function. One can rewrite the Majorana zero mode in the complex fermion basis $f_0$ as $\gamma_0=f_0+f_0^{\dagger}$. In terms of these $f$ operators, the coupling (\ref{MajoranaCoupling}) between the Majorana zero mode and the spin-polarized wire can be written as
\begin{equation}
H_{\Gamma1}=i\frac{\Gamma}{2}\int dx\, \delta(x)(f_{0}+f^{\dagger}_{0})(c_{x}^{\dagger}+c_{x}).
\label{NanowireMajoranaCoupling2}
\end{equation}
Because this coupling breaks the translation invariance of the unperturbed nanowire, we determine the nature of the pairing induced in the nanowire by calculating the Nambu Green's function in real space. We first rewrite the action in the Nambu spinor basis $\Psi^{\dagger}(x)=(c^{\dagger}_x,c_x,f^{\dagger}_0,f_0)$ as
$H=\frac{1}{2}\int dx \Psi^{\dagger}(x)\hat{\mathcal{H}}(x)\Psi(x)$, where the Bogoliubov-de Gennes (BdG) Hamiltonian matrix $\hat{\mathcal{H}}(x)$ can be split into two parts as $\hat{\mathcal{H}}(x)=\hat{\mathcal{H}}_{w}(x)+\hat{\mathcal{H}}_{\Gamma1}(x)$. The matrix $\hat{\mathcal{H}}_{w}(x)$ corresponds to the unperturbed BdG Hamiltonian for the spinless wire, which can be written as
\begin{eqnarray}
\hat{\mathcal{H}}_{w}(x)=
\begin{pmatrix}
    -\frac{\partial_x^2}{2m}-\varepsilon_F & 0 & 0 & 0 \\
    0 & \frac{\partial_x^2}{2m}+\varepsilon_F & 0 & 0 \\
    0 & 0 & 0 & 0 \\
    0 & 0 & 0 & 0
\end{pmatrix}.
\label{Hw}
\end{eqnarray}
From this Hamiltonian, we can calculate the Matsubara Green's function $G_0(x-y,i\omega_n)$ of the spinless nanowire before we turn on the coupling between the Majorana mode and the spinless wire. The unperturbed Green's function satisfies
\begin{eqnarray}
(i\omega_n-\hat{\mathcal{H}}_{w}(x))&&G_0(x-y,i\omega_n)\nonumber\\
&&=\begin{pmatrix}
    \delta(x-y) & 0 & 0 & 0 \\
    0 & \delta(x-y) & 0 & 0 \\
    0 & 0 & 1 & 0 \\
    0 & 0 & 0 & 1
\end{pmatrix},
\label{GreensFunctionEquation}
\end{eqnarray}
where $\omega_n$ is a fermionic Matsubara frequency. The unperturbed Green's function is therefore given by
\begin{align}
&G_0(x-y,i\omega_n)=\nonumber\\
&\begin{pmatrix}
    g_0(x-y,i\omega_n) & 0 & 0 & 0 \\
    0 & h_0(x-y,i\omega_n) & 0 & 0 \\
    0 & 0 & 1/(i\omega_n) & 0 \\
    0 & 0 & 0 & 1/(i\omega_n)
\end{pmatrix},
\label{GreensFunction}
\end{align}
where we use $g_0(x-y,i\omega_n)$ to denote the free electron propagator and $h_0(x-y,i\omega_n)$ to denote the free hole propagator. The free electron propagator in real space is given by Fourier transformation,
\begin{eqnarray}
g_0(x-y,i\omega_n)=\int \frac{dp}{2\pi}\frac{e^{ip(x-y)}}{i\omega_n-p^2/(2m)+\varepsilon_F}.
\label{ElectronPropagators}
\end{eqnarray}
The integrand has two poles in the complex $p$ plane given by $\pm p_0$ where $p_0=Je^{is\theta/2}$, and we define $\theta=\tan^{-1}(|\omega_n|/\varepsilon_F)$, $s=\text{sgn}(\omega_n)$, and $J=\sqrt{2m}(\varepsilon_F^2+\omega_n^2)^{1/4}$.

Performing the integral with the help of the residue theorem, we obtain the free electron propagator in real space as
\begin{eqnarray}
g_0(x-y,i\omega_n)=-\frac{mi}{p_0}se^{isp_0|x-y|},
\label{ElectronPropagators2}
\end{eqnarray}
and the free hole propagator as
\begin{eqnarray}
h_0(x-y,i\omega_n)&=&\int \frac{dp}{2\pi}\frac{e^{ip(x-y)}}{i\omega_n-[-p^2/(2m)+\varepsilon_F]}\nonumber\\
&=&-\frac{mi}{p_0^{*}}se^{-isp_0^{*}|x-y|},
\label{HolePropagators}
\end{eqnarray}
where $p_0^{*}=Je^{-is\theta/2}$ is the complex conjugate of $p_0$. Changing the sign of the Matsubara frequency $\omega_n$ corresponds to changing $p_0$ to $p_0^{*}$. (This is sensible as changing the sign of the Matsubara frequency $\omega_n$ results in the pole of the electron propagator $p_0$ becoming the pole of the hole propagator $p_0^*$.)

Now, we turn on the coupling between the Majorana mode and the spinless wire via the coupling Hamiltonian $\hat{\mathcal{H}}_{\Gamma1}(x)=M\delta(x)$ where the matrix $M$ is defined as
\begin{equation}
M=\frac{\Gamma}{2}
\begin{pmatrix}
    0 & 0 & -i & -i \\
    0 & 0 & -i & -i \\
    i & i & 0  & 0 \\
    i & i & 0  & 0
\end{pmatrix}.
\label{HGamma}
\end{equation}
The full Green's function $G(x,x',i\omega_n)$ in the presence of the coupling is given by the solution of the Dyson equation,
\begin{align}
G(x,y,i\omega_n)&=G_0(x-y,i\omega_n)+\int_{-\infty}^\infty dx' \,G_0(x-x',i\omega_n)\nonumber\\
&\hspace{30mm}\times\hat{\mathcal{H}}_{\Gamma1}(x')G(x'-y,i\omega_n)\nonumber\\
&=G_0(x-y,i\omega_n)+ G_0(x,i\omega_n)MG(0,y,i\omega_n).
\label{DysonEquation}
\end{align}
Setting $x=0$, we get
\begin{align}
G(0,y,i\omega_n)=G_0(-y,i\omega_n)+ G_0(0,i\omega_n)MG(0,y,i\omega_n),
\end{align}
which allows us to solve for $G(0,y,i\omega_n)$ as
\begin{align}
G(0,y,i\omega_n)=(1-G_0(0,i\omega_n)M)^{-1}G_0(-y,i\omega_n).
\label{Gy}
\end{align}
Substituting Eq.~(\ref{Gy}) back into the Dyson equation (\ref{DysonEquation}), we get the full Green's function as
\begin{align}
G(x,y,i\omega_n)&=G_0(x-y,i\omega_n)+G_0(x,i\omega_n)M\nonumber\\
&\hspace{5mm}\times[1-G_0(0,i\omega_n)M]^{-1}G_0(-y,i\omega_n).
\label{DysonEquation2}
\end{align}
The $(2,1)$ component of the full Green's function matrix $G(x,y,i\omega_n)$ gives the induced pairing correlator in the spinless nanowire,
\begin{align}
\langle c^{\dagger}_x(i\omega_n)c^{\dagger}_y(-i\omega_n)\rangle&=\frac{\Gamma^2h_0(x,i\omega_n)g_0(-y,i\omega_n)}{2i\omega_n-[g_0(0,i\omega_n)+h_0(0,i\omega_n)]\Gamma^2}\nonumber\\
&=\frac{-\Gamma^2m^2e^{-isp_0^{*}|x|}e^{isp_0|y|}}{2i\omega_n |p_0|^2+ism\Gamma^2(p_0+p_0^{*})}.
\label{PairingCorrelatorAPP}
\end{align}
Since changing the sign of the frequency $\omega_n$ corresponds to interchanging $p_0\leftrightarrow p_0^*$ and changing the sign of $s$, Eq.~(\ref{PairingCorrelatorAPP}) shows that the local pairing correlator $\langle c^{\dagger}_x(i\omega_n)c^{\dagger}_x(-i\omega_n)\rangle$ (i.e., with $y=x$) is an odd function of $\omega_n$. Therefore odd-frequency superconductivity is induced in the nanowire {\it even with a single Majorana zero mode}. This remarkable fact is at the origin of the robustness of the odd-frequency superconducting state induced by the method we propose: because odd-frequency pairing is induced in a local manner, it is largely insensitive to the ways in which translation symmetry is preserved/broken. In other words, since the induced odd-frequency superconductivity is $s$-wave, it is expected to be robust against translation symmetry breaking. In fact, pairing at the coupling site ($x=0$) remains odd in frequency even if we break translation symmetry in the wire Hamiltonian (\ref{SpinlessWire}). In that case the free electron and hole propagators are functions of $x$ and $y$ separately, i.e., $g_0(x-y,i\omega_n)$ becomes $g_0(x,y,i\omega_n)$ and likewise for $h_0$. For $x=y=0$, Eq.~(\ref{PairingCorrelatorAPP}) thus becomes
\begin{align}
&\langle c_0^\dag(i\omega_n)c_0^\dag(-i\omega_n)\rangle\nonumber\\
&\hspace{5mm}=-\frac{\Gamma^2 g_0(0,0,-i\omega_n)g_0(0,0,i\omega_n)}{2i\omega_n-[g_0(0,0,i\omega_n)-g_0(0,0,-i\omega_n)]\Gamma^2},
\end{align}
which is manifestly odd in $\omega_n$, using the fact that the hole and electron propagators are related by $h_0(x,y,i\omega_n)=-g_0(x,y,-i\omega_n)$.

In the low-frequency, weak-coupling limit $|\omega_n|\ll\Gamma\ll \varepsilon_F$, Eq.~(\ref{PairingCorrelatorAPP}) reduces to Eq.~(\ref{PairingCorrelatorApprox}) in the main text. The decay length for the pairing correlator depends on the frequency $\omega_n$ and diverges as $\omega_n\rightarrow 0$. To engineer uniform odd-frequency superconductivity, we can estimate a lower bound for the decay length by setting $\omega_n$ to $\varepsilon_F$, which gives a decay length $\sim v_F/\varepsilon_F\sim\lambda_F$ on the order of the Fermi wavelength. In the low-frequency limit $|\omega_n|\ll\varepsilon_F$, this decay length is in fact much greater than the Fermi wavelength. In other words, if the distance between Majorana modes is smaller than the Fermi wavelength $\lambda_F$, we effectively engineer uniform odd-frequency pairing in the wire. In the next section, we show this more explicitly, and also discuss the fate of odd-frequency pairing when the separation of the Majorana modes is larger than the Fermi wavelength.

\section{Continuum nanowire coupled to a periodic array of Majorana zero modes}
\label{sec:continuum}

In the main text we considered a simple model where the spin-polarized nanowire is effectively described by a tight-binding model, and the Majorana zero modes couple to every lattice site of this tight-binding model. Here we consider a more general case where the Majorana modes are coupled to a continuous nanowire. This introduces an additional parameter in the model, which is the ratio of the spacing $a$ between the Majorana modes to the Fermi wavelength $\lambda_F$ in the nanowire. As will be seen, the lattice model considered in the main text corresponds to the $a\ll\lambda_F$ limit in the continuous model, which is technically simpler. However, odd-frequency superconductivity is in fact obtained also away from this limit, as we will demonstrate here.

The nanowire is described by the Hamiltonian (\ref{SpinlessWire}) as previously, and coupled to a discrete array of Majorana zero modes (Fig.~\ref{MajoranaNanoWire}). The coupling to the Majorana modes is described by the Hamiltonian $H_\Gamma$ in Eq.~(\ref{NanowireMajoranaCoupling3}), and the hybridization between the Majorana modes at opposite ends of each superconducting wire is given by
\begin{eqnarray}
H_\delta=2\delta\int dx\sum_{n}\delta(x-na)(f_x^\dag f_x-\textstyle\frac{1}{2}).
\end{eqnarray}
The combined Hamiltonian $H_w+H_\Gamma+H_\delta$ then fully contains the gapless fermion modes propagating between the coupling sites in the nanowire.

\begin{figure}[t]
\includegraphics[width=8.5cm]{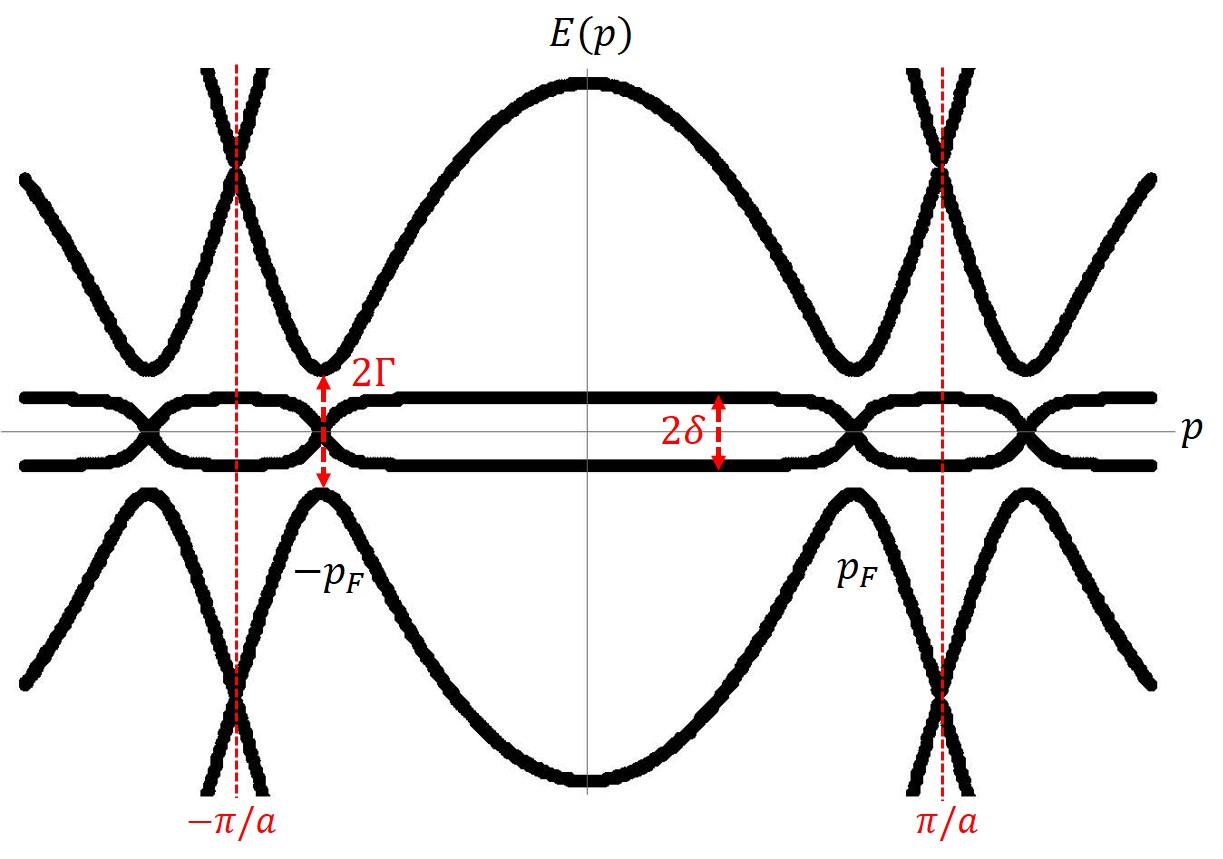}
\caption{Numerically calculated energy spectrum for a continuous spinless nanowire coupled to a periodic array of Majorana modes. As long as the Fermi points $\pm p_F$ are not too close to the Brillouin zone boundary $\pm\pi/a$, the low-energy spectrum is qualitatively the same as the one obtained from the lattice model described in the main text [compare with Fig.~\ref{BandStructureDOS}(a) in the main text].}
\label{EnergySpectrumPeriodicMZM}
\end{figure}

As mentioned in the main text, the lattice structure of the Majorana zero modes results in a periodic quasiparticle band structure with Brillouin zone boundary at $p=\pm\pi/a$. One can determine this band structure by diagonalizing the full Hamiltonian $H_w+H_{\Gamma}+H_\delta$ numerically~\footnote{In practice, for the numerical calculation we restrict the sum over $n$ to $-n_\text{max}\leq n\leq n_\text{max}$ where $n_\text{max}$ is increased until the spectrum converges.}, which gives the BdG spectrum shown in Fig.~\ref{EnergySpectrumPeriodicMZM}. This illustrates that if the Fermi points $\pm p_F$ are sufficiently far from the Brillouin zone boundary $\pm\pi/a$ (i.e., if $a$ is sufficiently small compared to $\lambda_F$), the low-energy spectrum is the same as obtained with the lattice model described in the main text [compare with Fig.~\ref{BandStructureDOS}(a)]. One can understand this limit as corresponding to the spinless nanowire uniformly coupling to a ``smeared'' Majorana mode, inducing spatially uniform odd-frequency superconductivity in the wire.

When the separation of the Majorana modes is larger than the Fermi wavelength, the induced odd-frequency pairing potential is no longer spatially uniform. The effect on the low-energy spectrum of band folding into the first Brillouin zone is more pronounced in this case, due to the coupling $(f_{p}+f^{\dagger}_{-p})(c_{p-2\pi n/a}^{\dagger}+c_{-p+2\pi n/a})$ between different momenta. In principle, one can simply calculate the Nambu Green's function of the full Hamiltonian $H_w+H_{\Gamma}+H_\delta$ and determine whether the induced pairing is odd-frequency. However, this is an infinite-dimensional problem if we take all the couplings $(f_{p}+f^{\dagger}_{-p})(c_{p-2\pi n/a}^{\dagger}+c_{-p+2\pi n/a})$, $n\in\mathbb{Z}$ into account. Due to the fact that we are only interested in low-energy properties, we can truncate the sum over $n$ in Eq.~(\ref{NanowireMajoranaCoupling3}) at a particular value of $n$ such that the condition $|\xi_{p_F\pm2n\pi/a}-\xi_{p_F}|\gg\Gamma$ is satisfied, where $\xi_{p_F}$ is the energy of the unperturbed wire at $p_F$ (in the main text our convention is such that $\xi_{p_F}=0$ by definition). In other words, since the Majorana modes couple to the bands at the Fermi energy with coupling strengh $\Gamma$, for small enough $\Gamma$ we can discard the high-energy bands and focus on the bands which are within $\Gamma$ of the Fermi energy $\xi_{p_F}$. With this truncation scheme, we are effectively projecting the Hamiltonian onto the low-energy subspace. This allows us to rewrite the Hamiltonian as a finite-dimensional matrix in momentum space, and use it to calculate the Nambu Green's function for the purposes of demonstrating odd-frequency pairing.

As an example, we demonstrate odd-frequency pairing in the case that the Majorana separation $a$ is larger than the Fermi wavelength $\lambda_F$, but small enough that only one folded band with energy $\xi_{p\pm2\pi/a}$ is coupled to the unfolded band $\xi_{p}$ by the Majorana coupling $H_{\Gamma}$ near the Fermi energy (i.e., $|\xi_{p_F\pm2\pi/a}-\xi_{p_F}|\gg\Gamma$). In this case, we can effectively project the Hamiltonian onto the three relevant bands $\xi_{p\pm2\pi/a}$ and $\xi_{p}$ at low energy. This allows us to write down a truncated coupling Hamiltonian at low energy as
\begin{align}
H_{\Gamma}=i\frac{\Gamma}{2}\sum_{n=-1}^{1}\int_{-\infty}^{\infty} \frac{dp}{2\pi} (f_{p}+f^{\dagger}_{-p})(c_{p-2\pi n/a}^{\dagger}+c_{-p+2\pi n/a}).
\end{align}
This reduces the Hamiltonian to a finite-dimensional matrix, which can be expressed explicitly in the basis $\Psi_{\text{trun}}^{\dagger}(p)=(c^{\dagger}_{p_-},c_{-p_-},c^{\dagger}_p,c_{-p},f^{\dagger}_p,f_{-p},c^{\dagger}_{p_+},c_{-p_+})$ where $p_{\pm}\equiv p\pm\frac{2\pi}{a}$. The truncated Hamiltonian matrix is
\begin{align}
&\mathcal{H}_{\text{trun}}(p)=\nonumber\\
&\hspace{5mm}\left(
\begin{array}{cccccccc}
 \xi_{p_-} & 0 & 0 & 0 & -\frac{i \Gamma }{2}  & -\frac{i \Gamma }{2}  & 0 & 0 \\
 0 & -\xi_{-p_-} & 0 & 0 & -\frac{i \Gamma }{2}  & -\frac{i \Gamma }{2}  & 0 & 0 \\
 0 & 0 & \xi_p  & 0 & -\frac{i \Gamma }{2}  & -\frac{i \Gamma }{2}  & 0 & 0 \\
 0 & 0 & 0 & -\xi_{-p}  & -\frac{i \Gamma }{2}  & -\frac{i \Gamma }{2}  & 0 & 0 \\
 \frac{i \Gamma }{2} & \frac{i \Gamma }{2} & \frac{i \Gamma }{2} & \frac{i \Gamma }{2} & \delta  & 0 & \frac{i \Gamma }{2} & \frac{i \Gamma }{2} \\
 \frac{i \Gamma }{2} & \frac{i \Gamma }{2} & \frac{i \Gamma }{2} & \frac{i \Gamma }{2} & 0 & -\delta  & \frac{i \Gamma }{2} & \frac{i \Gamma }{2} \\
 0 & 0 & 0 & 0 & -\frac{i \Gamma }{2}  & -\frac{i \Gamma }{2}  & \xi_{p_+} & 0 \\
 0 & 0 & 0 & 0 & -\frac{i \Gamma }{2}  & -\frac{i \Gamma }{2}  & 0 & -\xi_{-p_+} \\
\end{array}
\right).
\label{TruncatedHamiltonian}
\end{align}
From the Nambu Green's function $(i\omega_n-\mathcal{H}_{\text{trun}}(p))^{-1}$, we obtain the induced Gor'kov function in the spinless nanowire as
\begin{align}
&\langle c^{\dagger}_{-p}(-i\omega_n)c^{\dagger}_{p}(i\omega_n)\rangle
\nonumber\\
&\hspace{10mm}=\frac{i\omega_n\Gamma^2}{2}\frac{(\omega_n^2+\xi_{p_-}^2)(\omega_n^2+\xi_{p_+}^2)}{A+B\omega_n^2+C\omega_n^4+D\omega_n^6+\omega_n^8}.
\label{reduceOddSC}
\end{align}
Here we assume the inversion symmetry in the band structure $\xi_{-p}=\xi_{p}$, and the coefficients in the denominator of Eq.~(\ref{reduceOddSC}) are given by
\begin{align}
A&=\delta ^2\xi_{p_-}^2\xi_p^2\xi_{p_+}^2,\nonumber\\
B&=\xi_{p_-}^2\xi_p^2\xi_{p_+}^2+(\Gamma^2+\delta^2)(\xi_{p_-}^2\xi_{p_+}^2+\xi_{p_-}^2\xi_p^2+\xi_{p_+}^2\xi_p^2),\nonumber\\
C&=\xi_{p_-}^2\xi_{p_+}^2+\xi_{p_-}^2\xi_p^2+\xi_{p_+}^2\xi_p^2\nonumber\\
&\hspace{5mm}+(2 \Gamma ^2+\delta ^2)(\xi_{p_-}^2+\xi_p^2+\xi_{p_+}^2), \nonumber\\
D&=\xi_{p_-}^2+\xi_p^2+\xi_{p_+}^2+3 \Gamma ^2+\delta ^2.
\end{align}
This shows explicitly that the pairing correlator $\langle c^{\dagger}_{-p}(-i\omega_n)c^{\dagger}_{p}(i\omega_n)\rangle$ is still an odd function of frequency.

These calculations thus demonstrate that whether one chooses the nanowire to be continuous or discrete only affects the detailed form of the bands, not our fundamental conclusion that pure odd-frequency pairing is induced in such systems.

\section{Robustness of odd-frequency pairing against coupling disorder}

\label{sec:disorder}
\begin{figure}
\includegraphics[width=8.5cm]{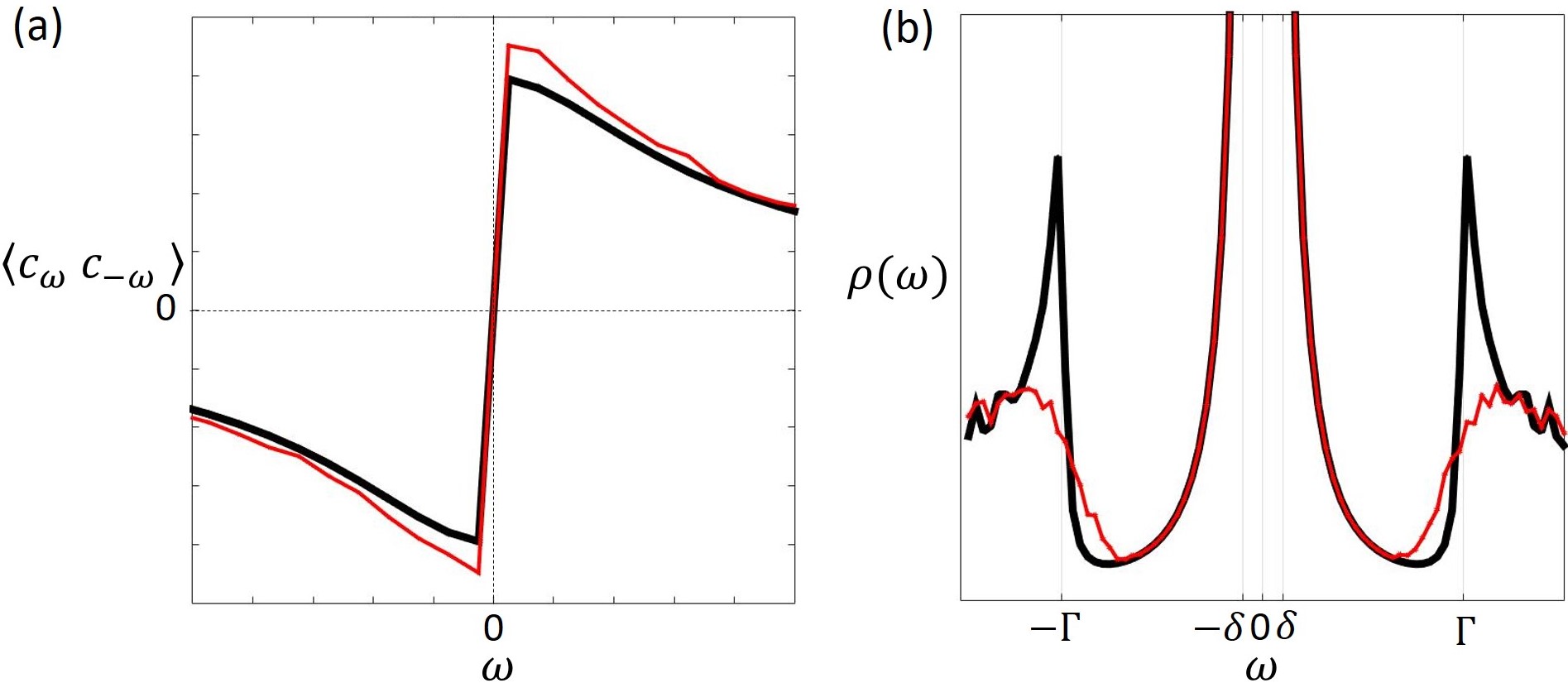}
\caption{(a) Disorder-averaged on-site pairing correlator as a function of frequency, (b) density of states as a function of energy. Red and black curves correspond to the case with and without disorder, respectively. We average over 20 configurations of disorder in a sample of 400 lattice sites. We take $\delta/\Gamma=0.1$.}
\label{Disorder}
\end{figure}
In this section we present evidence that the induced odd-frequency superconductivity in our setup is robust in an average sense against disorder in the coupling between Majorana modes and the nanowire. As explained in the main text, our original setup consists of a periodic array of $N$ Majorana zero modes, such that the $N$ coupling sites are not random but evenly spaced. If we couple the Majorana modes to every few lattice sites instead of to every site, the unit cell is effectively enlarged, leading to a reduced Brillouin zone and a folded BdG spectrum (see, e.g., Fig.~\ref{EnergySpectrumPeriodicMZM}). As discussed in Appendix~\ref{sec:continuum}, odd-frequency superconductivity does still appear in such circumstances [see Eq.~(\ref{reduceOddSC})]. In this section, we further demonstrate that odd-frequency pairing is robust against disorder by considering a random distribution of the Majorana couplings $\Gamma_i$. In restrospect, this fact might not be surprising because the odd-frequency superconductor we discuss in the main text is $s$-wave, which is expected to be robust against disorder.

To break translation invariance and model the effect of disorder, we imagine that the coupling strength between Majorana modes and the spinless wire varies from site to site. To do this, we Fourier transform the coupling Hamiltonian (\ref{NanowireMajoranaCouplingTrun}) to real space and consider a spatially varying coupling $\Gamma_i=\Gamma+\delta\Gamma_i$,
\begin{equation}
H_{\Gamma}=i\sum_{i=1}^{N} \frac{\Gamma_i}{2}\gamma_{i,a}(c_{i}^{\dagger}+c_{i}),
\label{NanowireMajoranaCoupling}
\end{equation}
where $\Gamma$ is the average coupling and $\delta\Gamma_i$ is the variation of the coupling from site to site. In our calculations the disorder strength in the coupling $\delta\Gamma_i$ is tuned to be the same as the average coupling $\Gamma$. In other words, the on-site Majorana coupling strength $\Gamma_i$ varies randomly between $0$ and $2\Gamma$ according to a uniform distribution. With this coupling, we can calculate the Nambu Green's function as $\mathcal{G}(i\omega_{n})=(i\omega_n-\mathcal{H})^{-1}$, where $\mathcal{H}$ is the total Hamiltonian including the wire Hamiltonian $H_w$, the tunnel coupling $H_\delta$, and the coupling of Majorana modes to the wire $H_{\Gamma}$ in real space. From the matrix elements of the Green's function, we can extract the on-site pairing correlator $\langle c_i(-\omega_n)c_i(\omega_n)\rangle$. Because translation symmetry is broken by disorder, the on-site pairing correlator $\langle c_i(-\omega_n)c_i(\omega_n)\rangle$ depends on the lattice site $i$ and the detailed configuration of the disorder, but one can take into account the average effect of disorder on pairing by defining a disorder-averaged pairing correlator. The latter is spatially uniform, but still depends on the Matsubara frequency $\omega_n$. In Fig.~\ref{Disorder}(a), we plot the disorder-averaged pairing correlator with (red curve) and without (black curve) disorder. As one can see, the pairing potential is odd in frequency even in the presence of disorder.  Similar to dirty $s$-wave superconductors, the disorder-averaged density of states exhibits smeared coherence peaks [Fig.~\ref{Disorder}(b)], but is otherwise qualitatively similar to the clean case (see Fig.~\ref{MajoranaNanoWire}(b) in the main text).

\section{Electromagnetic response: general formalism}
\label{APP:ElectromagneticResponseGeneralFormalism}

In this and the following two appendices we provide a detailed derivation of the electromagnetic response of our system. We consider a 1D superconductor with periodic boundary conditions, i.e., a superconducting ring. In second quantization, the action of the odd-frequency superconducting ring in imaginary time can be expressed in the Nambu basis $\Psi(p)=(c_p,c^{\dagger}_{-p},f_p,f^{\dagger}_{-p})^T$ as
\begin{align}
S[\Psi,\Psi^\dag]=-\frac{1}{2}T\sum_{ip_n}\int\frac{dp}{2\pi}\Psi^{\dagger}(p,ip_n) \c{G}^{-1}(p,ip_n)\Psi(p,ip_n),
\label{APP:action1}
\end{align}
at temperature $T$, where the imaginary-time Green's function $\c{G}$ is defined as
\begin{align}
\c{G}^{-1}(p,ip_n)&=ip_n-\mathcal{H}(p)\nonumber\\
&=\begin{pmatrix}
    ip_n-\xi_p & 0 & i\Gamma/2 & i\Gamma/2 \\
   0 & ip_n+\xi_{-p} & i\Gamma/2 & i\Gamma/2 \\
    -i\Gamma/2 & -i\Gamma/2 & ip_n-\delta & 0 \\
    -i\Gamma/2 & -i\Gamma/2 & 0 & ip_n+\delta
\end{pmatrix},
\label{APP:GreenFunction}
\end{align}
where $\xi_p$ is the energy-momentum dispersion of the spin-polarized wire measured with respect to the Fermi energy, and $p\equiv p_x$. We assume that the spin-polarized wire preserves inversion symmetry, such that $\xi_{-p}=\xi_p$. This is required to maintain a sharp distinction between the two possible forms of pairing in a 1D spinless system: even-frequency odd-parity, or odd-frequency even-parity. As a result, the velocity $d\xi_p/dp$ is odd in $p$ and the inverse effective mass $d^2\xi_p/dp^2$ is even in $p$.

To determine the Meissner response, we couple the ring to a vector potential $A_x$. The action in the presence of a vector potential is~\cite{RomanGaugeInvariant}
\begin{align}
S[\Psi,\Psi^\dag,A_x]&=-\frac{1}{2}\sum_{\tilde{p}}\Psi^{\dagger}(\tilde{p}) \c{G}^{-1}(\tilde{p})\Psi(\tilde{p})\nonumber\\
&\hspace{5mm}-\frac{1}{2}\sum_{\tilde{p},\tilde{q}}\Psi^\dag(\tilde{p}-\tilde{q}/2)J(\tilde{p},\tilde{q})\Psi(\tilde{p}+\tilde{q}/2)\nonumber\\
&\hspace{5mm}+\frac{1}{2}\sum_{\tilde{p},\tilde{q}}\Psi^\dag(\tilde{p})M(\tilde{p},\tilde{q})\Psi(\tilde{p}),
\label{APP:action}
\end{align}
where we use the simplified space-time notation $\tilde{p}=(p,ip_n)$ and $\sum_{\tilde{p}}\equiv T\sum_{ip_n}\int\frac{dp}{2\pi}$. The matrices $J$ and $M$ are defined as
\begin{eqnarray}
J(\tilde{p},\tilde{q})&=&e\frac{d\xi_{p}}{dp}\mathbb{I}A_x(\tilde{q}),\\
M(\tilde{p},\tilde{q})&=&\frac{e^2}{2}\frac{d^2\xi_p}{dp^2}\Sigma_zA_x(-\tilde{q})A_x(\tilde{q}),
\label{APP:Jx}
\end{eqnarray}
where we define
\begin{align}\label{APP:matrices}
\mathbb{I}=\begin{pmatrix}
    1 & 0 & 0 & 0 \\
    0 & 1 & 0 & 0 \\
    0 & 0 & 0 & 0 \\
    0 & 0 & 0 & 0
\end{pmatrix},\hspace{5mm}
\Sigma_z=\begin{pmatrix}
    1 & 0 & 0 & 0 \\
    0 & -1& 0 & 0 \\
    0 & 0 & 0 & 0 \\
    0 & 0 & 0 & 0
\end{pmatrix}.
\end{align}
In Eq.~(\ref{APP:Jx}) we have set $\tilde{q}=0$ in the vertices $d\xi_p/dp$ and $d^2\xi_p/dp^2$. Furthermore, strictly speaking the momentum dependence of the last term in Eq.~(\ref{APP:action}) only appears after performing the path integral over the fermions~\cite{RomanGaugeInvariant}. We can integrate out the fermion fields to obtain an effective action for the vector potential, which we expand to second order in the latter,
\begin{align}
S_{\text{eff}}[A_x]&=-\ln \Pf\left(-\c{G}^{-1}-J+M\right)\nonumber\\
&=-\frac{1}{2}\Tr\ln\left(-\c{G}^{-1}\right)-\frac{1}{2}\Tr\c{G}J+\frac{1}{2}\Tr\c{G}M\nonumber\\
&\hspace{5mm}+\frac{1}{4}\Tr\c{G}J\c{G}J+\mathcal{O}(A_x^3),
\label{APP:EffectiveAction}
\end{align}
where the trace is to be understood in the functional sense (i.e., trace over Nambu spinor indices, sum over Matsubara frequencies, and integral over momentum). Because the Nambu spinor satisfies the (Majorana) condition $\Psi^\dag(x)=\Psi^T(x)C$ where $C=\sigma_x\oplus\sigma_x$ is a charge-conjugation matrix, $\Psi^\dag$ and $\Psi$ are not independent variables in the functional integral and one obtains a Pfaffian instead of a determinant in Eq.~(\ref{APP:EffectiveAction}) (see, e.g., the Appendix of Ref.~\onlinecite{ariad2015}). In linear response theory, the response current is linearly proportional to the gauge field $A_x$. Because the electric current $j_x$ is given by the functional derivative of the effective action $j_x=-\delta S_{\text{eff}}/\delta A_x$, we focus on the terms quadratic in $A_x$ in the effective action $S_{\text{eff}}$. There are two such terms: the diamagnetic term $\Tr\c{G}M$ and the paramagnetic term $\Tr\c{G}J\c{G}J$.

\section{Diamagnetic response\label{APP:DiamagneticCurrent}}

In a ring geometry, the $x$ coordinate can be viewed as a coordinate along the circumference of the ring with radius $R$, such that $x=R\theta$ where $\theta$ is the polar angle. If we consider the response to an applied magnetic flux $\Phi$ threading the ring, the corresponding vector potential is a constant, $A_x=\Phi/2\pi R$. The diamagnetic part of the effective action (\ref{APP:EffectiveAction}) is
\begin{align}\label{supp:Sdia}
&S_\text{dia}[A_x]=\frac{1}{2}\Tr\c{G}M\nonumber\\
&\hspace{5mm}=\frac{e^2A_x^2}{4}T\sum_{ip_n}\int\frac{dp}{2\pi}\frac{d^2\epsilon_p}{dp^2}\tr\left[\c{G}(p,ip_n)\Sigma_ze^{ip_n0^+\Sigma_z}\right],
\end{align}
where the convergence factor $e^{ip_n0^+\Sigma_z}$ is necessary~\cite{ariad2015} to ensure that the total electron density $n$ in the wire is correctly given by $n=\frac{1}{2}\Tr\c{G}\Sigma_z$ in the absence of coupling $\Gamma$ between the wire and the Majorana zero modes. By contrast with the functional trace $\Tr$, the lowercase trace symbol $\tr$ on the right-hand side of Eq.~(\ref{supp:Sdia}) denotes a trace over Nambu spinor indices only. We also neglect a factor of spacetime volume in the action coming from the fact that the vector potential considered is constant in time and uniform in space.

We can first verify that Eq.~(\ref{supp:Sdia}) gives the correct result for the diamagnetic response of the metallic wire in its normal state, when it is decoupled from the Majorana zero modes. Setting $\Gamma=0$ in Eq.~(\ref{APP:GreenFunction}), we obtain
\begin{align}\label{supp:GFDeltaZero}
\c{G}(p,ip_n)=\diag\left(\frac{1}{ip_n-\xi_p},\frac{1}{ip_n+\xi_p},\frac{1}{ip_n-\delta},\frac{1}{ip_n+\delta}\right),
\end{align}
using $\xi_{-p}=\xi_p$, and Eq.~(\ref{supp:Sdia}) becomes
\begin{align}
S_\text{dia}[A_x]&=\frac{e^2A_x^2}{4}\int\frac{dp}{2\pi}\frac{d^2\epsilon_p}{dp^2}T\sum_{ip_n}
\left(\frac{e^{ip_n0^+}}{ip_n-\xi_p}-\frac{e^{-ip_n0^+}}{ip_n+\xi_p}\right)\nonumber\\
&=\frac{e^2A_x^2}{2}\int\frac{dp}{2\pi}\frac{d^2\epsilon_p}{dp^2}n_F(\xi_p),
\end{align}
where $n_F(\xi)=(e^{\xi/T}+1)^{-1}$ is the Fermi-Dirac distribution and $\xi_p$ is the energy of single particle excitations. For $p$ near the Fermi momentum $\pm p_F$, we have $\xi_p\approx\pm v_F(p\mp p_F)$. In the effective mass approximation we have $d^2\xi_p/dp^2=1/m$, and we obtain the standard diamagnetic response
\begin{eqnarray}\label{supp:normaldia}
S_\text{dia}[A_x]&=&\frac{ne^2}{2m}A_x^2,\\
j_x^\text{dia}(\Gamma=0)&=&-\frac{\delta S_\text{dia}}{\delta A_x}=-\frac{ne^2}{m}A_x,
\end{eqnarray}
where $n=\int(dp/2\pi)n_F(\xi_p)$ is the electronic density of the metallic wire.

We now consider the diamagnetic response of the odd-frequency superconductor with $\Gamma\neq 0$. Once again in the effective mass approximation, we have
\begin{align}\label{supp:jdiabeforesum}
j_x^\text{dia}=-\frac{e^2A_x}{2m}T\sum_{ip_n}\int\frac{dp}{2\pi}\tr\left[\c{G}(p,ip_n)\Sigma_ze^{ip_n0^+\Sigma_z}\right],
\end{align}
where $\c{G}$ is now the full Green's function (\ref{APP:GreenFunction}) with $\Gamma\neq 0$. To perform the sum over Matsubara frequencies, it is most convenient to introduce the spectral function $\c{A}(p,\omega)$, in terms of which the Green's function is given by
\begin{align}\label{supp:defA}
\c{G}(p,ip_n)=\int\frac{d\omega}{2\pi}\frac{\c{A}(p,\omega)}{ip_n-\omega}.
\end{align}
For the quadratic Hamiltonian considered here, the spectral function can be written as
\begin{align}\label{supp:specfunc}
\c{A}(p,\omega)=2\pi\sum_A\Res_{z=E_A(p)}\c{G}(p,z)\delta\left(\omega-E_A(p)\right),
\end{align}
where $\c{G}(p,z)=(z-\c{H}(p))^{-1}$ and the sum is over the four Bogoliubov energy bands plotted in Fig.~\ref{BandStructureDOS}(a) of the main text: $E_1(p)$, $E_2(p)$, $E_3(p)=-E_1(p)$, and $E_4(p)=-E_2(p)$, with
\begin{eqnarray}\label{supp:bandstructure}
&&E_1(p)=\frac{1}{\sqrt{2}}\sqrt{\xi_p^2+\Gamma^2+\delta^2+Y},\\
&&E_2(p)=\frac{1}{\sqrt{2}}\sqrt{\xi_p^2+\Gamma^2+\delta^2-Y},
\end{eqnarray}
where $Y=\sqrt{(\xi_p^2+\Gamma^2)^2+2\delta^2(\Gamma^2-\xi_p^2)+\delta^4}$. The residues of the Green's function are defined as
\begin{align}
\Res_{z=E_A(p)}\c{G}(p,z)=\lim_{z\rightarrow E_A(p)}(z-E_A(p))\c{G}(p,z).
\end{align}
We obtain
\begin{align}
\c{A}(p,\omega)&=\frac{\pi}{E_1(E_1^2-E_2^2)}\left[\delta(\omega-E_1)-\delta(\omega+E_1)\right]\nonumber\\
&\hspace{5mm}\times\begin{pmatrix}
p(\omega)+L(\omega) & \Gamma^2\omega/2 & \cdots & \cdots \\
\Gamma^2\omega/2 & p(\omega)-L(\omega) & \cdots & \cdots \\
\cdots & \cdots & \cdots & \cdots \\
\cdots & \cdots & \cdots & \cdots
\end{pmatrix}\nonumber\\
&\hspace{5mm}+(E_1\leftrightarrow E_2),
\label{APP:SpectralFunction}
\end{align}
where we define
\begin{align}
p(\omega)=\omega^3-\left(\frac{\Gamma^2}{2}+\delta^2\right)\omega,\hspace{5mm}
L(\omega)=\xi_p(\omega^2-\delta^2).
\end{align}
It is not necessary to calculate the matrix elements denoted by $\cdots$ in Eq.~(\ref{APP:SpectralFunction}); they contribute nothing to the trace in Eq.~(\ref{supp:jdiabeforesum}) because of Eq.~(\ref{APP:matrices}).

Performing the sum over Matsubara frequencies in Eq.~(\ref{supp:jdiabeforesum}) using Eq.~(\ref{supp:specfunc}), we obtain
\begin{align}\label{APP:OddSCDiamagnetic}
j_x^\text{dia}&=-\frac{e^2A_x}{2m}\int\frac{dp}{2\pi}\frac{1}{E_1^2-E_2^2}\nonumber\\
&\hspace{10mm}\times\left(\frac{p(E_1)-L(E_1)+2L(E_1)n_F(E_1)}{E_1}\right.\nonumber\\
&\hspace{15mm}\left.-\frac{p(E_2)-L(E_2)+2L(E_2)n_F(E_2)}{E_2}\right).
\end{align}
Focusing on the zero-temperature limit, because $E_1$ and $E_2$ are positive we have $n_F(E_1)=n_F(E_2)=0$ and Eq.~(\ref{APP:OddSCDiamagnetic}) reduces to
\begin{eqnarray}\label{supp:jdia}
j_x^\text{dia}=&-&\frac{e^2A_x}{2m}\int\frac{dp}{2\pi}\frac{1}{E_1^2-E_2^2}\left(\frac{p(E_1)-L(E_1)}{E_1}\right.\nonumber\\
&-&\left.\frac{p(E_2)-L(E_2)}{E_2}\right).
\end{eqnarray}
We will evaluate this integral approximately in the limit $\delta\ll\Gamma\ll\varepsilon_F$. Let us define the change in diamagnetic current from its value at zero coupling $\Gamma=0$:
\begin{align}\label{supp:deltaj}
\delta j_x^\text{dia}(\Gamma)=j_x^\text{dia}(\Gamma)-j_x^\text{dia}(\Gamma=0).
\end{align}
In the limit $\delta\ll\Gamma\ll\varepsilon_F$, the changes in the bandstructure and wave functions due to a finite coupling $\Gamma$ are confined to a small interval in momentum space of order $\pm\Lambda$ around the Fermi points $\pm p_F$, where $\Lambda$ is of order $\sim\Gamma/v_F$. This can be thought of as the weak-pairing limit of BCS theory. Therefore we can obtain an approximate expression for (\ref{supp:deltaj}) by restricting the integral in (\ref{supp:jdia}) to this interval around the Fermi points. Because $\xi_p$ vanishes at the Fermi points, in this interval we can approximate $\xi_p\ll\Gamma$. Together with the condition $\delta\ll\Gamma$, we obtain the approximate expressions
\begin{align}\label{eq:endefined}
E_1(p)\approx\Gamma,\hspace{5mm}E_2(p)\approx\frac{\delta|\xi_p|}{\Gamma},
\end{align}
which further implies $E_1^2-E_2^2\approx\Gamma^2$, $p(E_1)-L(E_1)\approx\Gamma^3/2$, and $p(E_2)-L(E_2)\approx-\delta\Gamma|\xi_p|/2$. Substituting in Eq.~(\ref{supp:jdia}), we obtain
\begin{eqnarray}
\delta j_x^\text{dia}(\Gamma)&\approx&-\frac{e^2A_x}{2m}\left(\int_{-p_F-\Lambda}^{-p_F+\Lambda}\frac{dp}{2\pi}+\int_{p_F-\Lambda}^{p_F+\Lambda}\frac{dp}{2\pi}\right)\nonumber\\
&\approx&-\frac{\Lambda}{\pi}\frac{e^2A_x}{m}.
\end{eqnarray}
If we estimate the momentum-space cutoff as $\Lambda=\Gamma/v_F$ and use $n=p_F/\pi$ for the density and $\varepsilon_F=\frac{1}{2}v_Fp_F$ for the Fermi energy, we obtain
\begin{align}\label{supp:jdiaZeroT}
j_x^\text{dia}(\Gamma)&=j_x^\text{dia}(\Gamma=0 )+\delta j_x^\text{dia}(\Gamma)\nonumber\\
&\approx-n\left(1+\frac{\Gamma}{2\varepsilon_F}\right)\frac{e^2A_x}{m},
\end{align}
at zero temperature. Thus in the presence of the coupling $\Gamma$ between the spin-polarized wire and the Majorana zero modes, the diamagnetic response is simply that of the decoupled wire plus a small correction of order $\Gamma/\varepsilon_F\ll 1$. We further assume that superconducting phase fluctuations are suppressed and phase coherence is maintained among the wires, for example via the common ground (see Fig.~\ref{MajoranaNanoWire}) or a direct Josephson coupling that could be engineered between the wires. Such an inter-wire Josephson coupling will in general contribute an additional diamagnetic current. However, one can always build the Josephson junctions or the common ground such that they are far away from the spin-polarized wire. Since the induced paramagnetic current is localized on the spin-polarized ring, one can use a local probe to detect the nontrivial paramagnetic current in the odd-frequency superconductor without picking up a (potentially larger) diamagnetic contribution coming from the conventional Josephson effect between the wires.

\section{Paramagnetic response \label{APP:ParamagneticCurrent}}

The paramagnetic part of the effective action (\ref{APP:EffectiveAction}) is
\begin{align}
&S_\text{para}[A_x]=\frac{1}{4}\Tr\c{G}J\c{G}J\nonumber\\
&=T\sum_{iq_n}\int\frac{dq}{2\pi}Q_{xx}(q,iq_n)A_x(q,iq_n)A_x(-q,-iq_n),
\end{align}
where the current-current correlation function $Q_{xx}(q,iq_n)$ is
\begin{align}\label{supp:JJcorr}
&Q_{xx}(q,iq_n)=\frac{e^2}{4}T\sum_{ip_n}\int\frac{dp}{2\pi}\left(\frac{d\xi_p}{dp}\right)^2\nonumber\\
&\times\tr \left[\c{G}(p-q/2,ip_n-iq_n/2)\mathbb{I}\c{G}(p+q/2,ip_n+iq_n/2)\mathbb{I}\right],
\end{align}
where $\xi_p$ was defined below Eq.~\eqref{supp:GFDeltaZero}.
Using the spectral representation (\ref{supp:defA}) and the Matsubara sum
\begin{align}\label{supp:MatsubaraSum}
T\sum_{ip_n}\frac{1}{(ip_n-\frac{iq_n}{2}-\omega)(ip_n+\frac{iq_n}{2}-\tilde{\omega})}
=\frac{n_F(\omega)-n_F(\tilde{\omega})}{iq_n+\omega-\tilde{\omega}},
\end{align}
we obtain
\begin{align}
&Q_{xx}(q,iq_n)=\frac{e^2}{4}\int\frac{dp}{2\pi}\left(\frac{d\xi_p}{dp}\right)^2\int\frac{d\omega}{2\pi}\int\frac{d\tilde{\omega}}{2\pi}\nonumber\\
&\times\left(\frac{n_F(\omega)-n_F(\tilde{\omega})}{iq_n+\omega-\tilde{\omega}}\right)\tr\left[\c{A}(p-q/2,\omega)\mathbb{I}\c{A}(p+q/2,\tilde{\omega})\mathbb{I}\right].
\end{align}
For the Meissner response, we are interested in the static (thermodynamic) susceptibility $Q_{xx}(q)\equiv\lim_{iq_n\rightarrow 0}Q_{xx}(q,iq_n)$,
\begin{align}\label{supp:Qxxintegral}
&Q_{xx}(q)=\frac{e^2}{4}\int\frac{dp}{2\pi}\left(\frac{d\xi_p}{dp}\right)^2\int\frac{d\omega}{2\pi}\int\frac{d\tilde{\omega}}{2\pi}\nonumber\\
&\times\left(\frac{n_F(\omega)-n_F(\tilde{\omega})}{\omega-\tilde{\omega}}\right)\tr\left[\c{A}(p-q/2,\omega)\mathbb{I}\c{A}(p+q/2,\tilde{\omega})\mathbb{I}\right].
\end{align}
As in the previous section, in the $\delta\ll\Gamma\ll\varepsilon_F$ limit we can focus on changes to the paramagnetic response $\delta Q_{xx}(q)\equiv Q_{xx}(q)_\Gamma-Q_{xx}(q)_{\Gamma=0}$ due to a finite $\Gamma$, which arises from small intervals in momentum space of order $\pm\Gamma/v_F$ around the Fermi points. This is sufficient since in the $\Gamma\rightarrow 0$ limit the spin-polarized wire is decoupled from the Majorana zero modes, and its total superfluid response (diamagnetic plus paramagnetic) vanishes. In fact, one can check explicitly from Eq.~(\ref{supp:GFDeltaZero}) and (\ref{supp:JJcorr})-(\ref{supp:MatsubaraSum}) that the paramagnetic current $j_x^\text{para}=-\delta S_\text{para}/\delta A_x$ equals $ne^2A_x/m$ for $\Gamma=0$, which exactly cancels the $\Gamma=0$ diamagnetic current in Eq.~(\ref{supp:jdiaZeroT}). Going back to the $\Gamma\neq 0$ case, at low temperatures and in the long-wavelength limit $q\rightarrow 0$, the dominant contribution to (\ref{supp:Qxxintegral}) comes from states within $\sim T$ of the Fermi level. In the limit $T\ll\Gamma$, the only such states are the $\pm E_2(p)$ bands. We can thus obtain an approximate expression for the paramagnetic response in this limit by neglecting the contribution of the $\pm E_1(p)$ bands to the spectral function. In the limit $\delta\ll\Gamma\ll\varepsilon_F$ and near the Fermi points, the spectral function neglecting the contribution of the $\pm E_1(p)$ bands is given by
\begin{align}
&\c{A}(p,\omega)\approx\nonumber\\
&\frac{\pi}{2}\left[\delta(\omega-E_2(p))-\delta(\omega+E_2(p))\right]
\begin{pmatrix}
1 & -1 & \cdots & \cdots \\
-1 & 1 & \cdots & \cdots \\
\cdots & \cdots & \cdots & \cdots \\
\cdots & \cdots & \cdots & \cdots
\end{pmatrix},
\end{align}
and we obtain
\begin{align}
\delta Q_{xx}(q)\approx\frac{e^2v_F^2}{16}\sum_{s,\tilde{s}}\int_{|p\pm p_F|<\Lambda}\frac{dp}{2\pi}
\frac{n_F(sE_2)-n_F(\tilde{s}\tilde{E}_2)}{sE_2-\tilde{s}\tilde{E}_2},
\end{align}
where $s,\tilde{s}=\pm 1$ and we use the notation $E_2\equiv E_2(p-q/2)$, $\tilde{E}_2\equiv E_2(p+q/2)$.

In the long-wavelength limit $q\rightarrow 0$, the Fermi-Dirac distribution can be expanded in powers of $q$ and we have
\begin{align}\label{supp:dQxx}
&\delta Q_{xx}(0)\approx\nonumber\\
&\frac{e^2v_F^2}{8}\int_{|p\pm p_F|<\Lambda}\frac{dp}{2\pi}
\left[\frac{dn_F[E_2(p)]}{dE_2(p)}+\frac{n_F[E_2(p)]-\frac{1}{2}}{E_2(p)}\right],
\end{align}
where $E_2(p)$ is defined in Eq. \eqref{eq:endefined}.

One may then consider two limits, the $T\ll\delta$ and the $\delta\ll T$ limits. In the $T\ll\delta$ limit, we can take the zero-temperature limit of the derivative of the Fermi-Dirac distribution, which becomes a delta function. The first term of (\ref{supp:dQxx}) becomes
\begin{align}\label{supp:Tlldelta1}
\int_{|p\pm p_F|<\Lambda}\frac{dp}{2\pi}\frac{dn_F[E_2(p)]}{dE_2(p)}&\approx-\int_{|p\pm p_F|<\Lambda}\frac{dp}{2\pi}\delta[E_2(p)]\nonumber\\
&=-\frac{1}{\pi v_F}\frac{\Gamma}{\delta}.
\end{align}
In the last equality, we linearize the dispersion of the spin polarized nanowire as $\xi_p\approx\pm v_F(p\mp p_F)$ and use Eq.~(\ref{eq:endefined}) for the $E_2(p)$ inside the delta function. The second term of (\ref{supp:dQxx}) can be written as
\begin{align}
&\int_{|p\pm p_F|<\Lambda}\frac{dp}{2\pi}\frac{n_F[E_2(p)]-\frac{1}{2}}{E_2(p)}\nonumber\\
&\hspace{10mm}=-\frac{1}{\pi v_F}\frac{\Gamma}{\delta}\int_0^{\delta/T}\frac{dx}{x}\tanh\left(\frac{x}{2}\right).
\end{align}
In the limit $T\ll\delta$, the dominant contribution to this integral comes from the $x\gg 1$ region, where $\tanh(x/2)\approx 1$, and we obtain
\begin{align}\label{supp:Tlldelta2}
\int_{|p\pm p_F|<\Lambda}\frac{dp}{2\pi}\frac{n_F[E_2(p)]-\frac{1}{2}}{E_2(p)}\approx-\frac{1}{\pi v_F}\frac{\Gamma}{\delta}\ln\left(\frac{\delta}{T}\right).
\end{align}
Putting (\ref{supp:Tlldelta1}) and (\ref{supp:Tlldelta2}) together, we find
\begin{align}
\delta Q_{xx}(0)\approx-\frac{e^2v_F}{8\pi}\frac{\Gamma}{\delta}\left[1+\ln\left(\frac{\delta}{T}\right)\right],
\end{align}
in the $T\ll\delta$ limit.

In the $\delta\ll T$ limit, we can set $\delta=0$ rather than $T=0$ in the derivative of the Fermi-Dirac distribution $dn_F[E_2(p)]/dE_2(p)\approx-1/4T$. The Fermi-Dirac distribution in the second term of (\ref{supp:dQxx}) can be expanded to first order in its argument, since as $p$ ranges from $-\Lambda$ to $\Lambda$ the dimensionless argument $E/T$ of the Fermi-Dirac distribution $n_F(E)$ ranges from $-\delta/T$ to $\delta/T$ (assuming $\Lambda=\Gamma/v_F$), which is much less than one in the limit considered. Both terms in (\ref{supp:dQxx}) then contribute equally, and we obtain
\begin{align}
\delta Q_{xx}(0)\approx-\frac{e^2v_F}{8\pi}\frac{\Gamma}{T},
\end{align}
in the $\delta\ll T$ limit.

The change in paramagnetic current $\delta j_x^\text{para}(\Gamma)=j_x^\text{para}(\Gamma)-j_x^\text{para}(\Gamma=0)$ is thus given by
\begin{align}\label{supp:djpara}
\delta j_x^\text{para}(\Gamma)=\frac{ne^2A_x}{4m}\times\left\{
\begin{array}{cc}
\displaystyle\frac{\Gamma}{\delta}\left[1+\ln\left(\frac{\delta}{T}\right)\right], & T\ll\delta, \\
\displaystyle\frac{\Gamma}{T}, & \delta\ll T,
\end{array}
\right.
\end{align}
where we have used $v_F/\pi=p_F/m\pi=n/m$. Apart from the logarithmically divergent term in the $T\ll\delta$ limit, the two small energy scales $T$ and $\delta$ act as an infrared cutoff for each other.

\section{Superfluid density}\label{APP:SuperfluidDensity}

Defining the superfluid density $n_s$ via the London equation $j_x=-\frac{n_se^2}{m}A_x$ where $j_x=j_x^\text{dia}+j_x^\text{para}$ is the total current, we obtain (taking into account the fact that $n_s$ vanishes for $\Gamma=0$)
\begin{align}
\frac{n_s}{n}\approx \left\{
\begin{array}{cc}
\displaystyle \frac{\Gamma}{2\varepsilon_F}-\frac{\Gamma}{4\delta}\left[1+\ln\left(\frac{\delta}{T}\right)\right], & T\ll\delta, \\
\displaystyle \frac{\Gamma}{2\varepsilon_F}- \frac{\Gamma}{4T}, & \delta\ll T,
\end{array}
\right.
\end{align}
Because we consider the limit $T,\delta\ll\Gamma\ll\varepsilon_F$, the diamagnetic term is negligible in front of the paramagnetic term, and we obtain
\begin{align}
\frac{n_s}{n}\approx-\frac{\Gamma}{4T}\times\left\{
\begin{array}{cc}
\displaystyle\frac{T}{\delta}\left[1+\ln\left(\frac{\delta}{T}\right)\right], & T\ll\delta, \\
\displaystyle 1, & \delta\ll T,
\end{array}
\right.
\end{align}
The odd-frequency superconductor thus exhibits a negative superfluid density, indicating a paramagnetic Meissner effect.
%\bibliography{OddFrequencySC-ref}

\end{document}